# MODELING HEAT TRANSFER FROM QUENCH PROTECTION HEATERS TO SUPERCONDUCTING CABLES IN NB3SN MAGNETS

T. Salmi[*], D. Arbelaez, S. Caspi, H. Felice, S. Prestemon, LBNL, Berkeley, CA 94720, USA

G. Chlachidze, Fermi National Laboratory, IL, USA

H. H. J. ten Kate, University of Twente, Enschede, the Netherlands

*Abstract*

We use a recently developed quench protection heater modeling tool for an analysis of heater delays in superconducting high-field Nb$_3$Sn accelerator magnets. The results suggest that the calculated delays are consistent with experimental data, and show how the heater delay depends on the main heater design parameters.

## INTRODUCTION

The quench protection of the present-day high-field Nb$_3$Sn accelerator magnets is based on resistive protection heaters – typically stainless steel–polyimide laminates on the coil surfaces [1]. They bring large segments of the winding to a resistive state during quench, accelerating the magnet current decay and consequently reducing the hotspot temperature. The goal of the heater design is to provide a short heater delay, i.e. the time delay between heater activation and the heater induced quench, and quench a large fraction of the winding. Physical limitations come from the maximum heater voltage and temperature (typically 400 V and 350 K, respectively). The heater insulation thickness (typically between 0.025 and 0.100 mm) required for electrical integrity has a significant effect on the delay time.

In the magnets under development for the LHC HiLumi upgrade, whose length is of the order of 10 m, the heater delay should be in the order of 10 ms, and the heaters should cover at least 60-100% of the coil surface [2][3]. This has been obtained in shorter and/or lower energy R&D magnets (LARP LQ and HQ) [3], but now the increased coil surface area and also requirements for thicker heater insulation to guarantee electrical integrity (increase from 0.025 mm to 0.050-0.075 mm) bring new challenges. Also, LQ and HQ, which had heaters on both the coils inner and outer surfaces, showed that only the outer surface heaters are mechanically reliable. Therefore, significant optimization of the present technology is needed. An additional complexity comes from the need of heating stations for long magnets, making the geometry of the heater non-uniform along the magnet length and adding an additional degree of freedom to the heater design problem.

This paper summarizes a recently developed numerical modeling tool for simulating heat transfer between the heater and coil. The model accounts for the heater geometry and powering, the cable properties, magnetic field and the various insulation materials allowing the evaluation of the heater delay in different conditions. The model is first applied to the LARP HQ magnet [4]. First, the real heater geometry is simulated and the delays are compared with experimental data from [5]. Second, a parametric analysis is used to examine the impact of main heater design parameters on the quench delay. The model is then applied to simulate the protection heaters in the so-called 11 T dipole prototype, built within a CERN and FNAL joint R&D program [6], and the simulated delays are compared with experimental data. Understanding of the impact of the heater design on the quench delay is important for designing the protection for future magnets.

## COMPUTATIONAL MODEL

*Thermal model*

The heat transfer between the heater and the cable is simulated using a numerical two-dimensional heat conduction model, with joule heat generation in the stainless steel component to simulate heater powering. In this approximation the heat propagation between neighboring turns is neglected. At the present stage of development, current sharing between the strands and quench propagation due to Joule heating in the cable is also not simulated.

The two-dimensional heat equation describing the thermal propagation is

$$\gamma_m c_{p,m} \frac{\partial T}{\partial t} = \frac{\partial}{\partial y}\left(k_m \frac{\partial T}{\partial y}\right) + \frac{\partial}{\partial z}\left(k_m \frac{\partial T}{\partial z}\right) + f_{gen,ss}, \qquad (1)$$

where $T = T(z,y,t)$ is temperature (K), $c_{p,m} = c_{p,m}(T,B)$ is specific heat (J/K/kg), $\gamma_m$ is mass density (kg/m$^3$), $k_m = k_m(T,B)$ is thermal conductivity (W/K/m) of the material $m$ at the location $(z, y)$ at time $t$ (s) and $f_{gen,ss} = f_{gen,ss}(t,T)$ is the internal volumetric heat source applied only in stainless steel component. (W/m$^3$). It is defined using

$$f_{gen,ss}(t,T) = \rho_{ss}(T) J_{ss}^2(t), \qquad (2)$$

where $J_{ss}(t)$ is the heater current density (A/m$^2$) and $\rho_{ss}(T)$ is the stainless steel electrical resistivity ($\Omega$m), or using

$$f_{gen,ss}(t) = P_{PH}(0)/d_{ss}\, e^{-\frac{2t}{\tau}}, \qquad (3)$$

where $P_{PH}(0)$ (W/m$^2$) is the heater adiabatic peak power defined by dividing the heater power by the heating surface area [1], $d_{ss}$ (m) is the stainless steel thickness, and $\tau$ is a time constant of an exponential heater current decay.

---

A heater on the coil straight section typically has a periodical geometry (see Fig. 1). Due to the symmetry, each turn can be represented by modeling only half of the heater period, when adiabatic boundary conditions are assumed at the center and at the end of the period ($z = 0$, and at $z = PH\ period/2$). Figure 2 shows a case with one heating segment at the center of the heater period. The boundaries at the top and bottom of the system, i.e. at $y = 0$, or at $y = H$, are at fixed temperature, $T_{bath}$.

*Material properties and magnetic field*

The various insulation layers as well as the cable and heater dimensions are taken into account using regions of different material properties. The different layers are assumed in perfect thermal contact. The layers dimensions and materials are an input parameter.

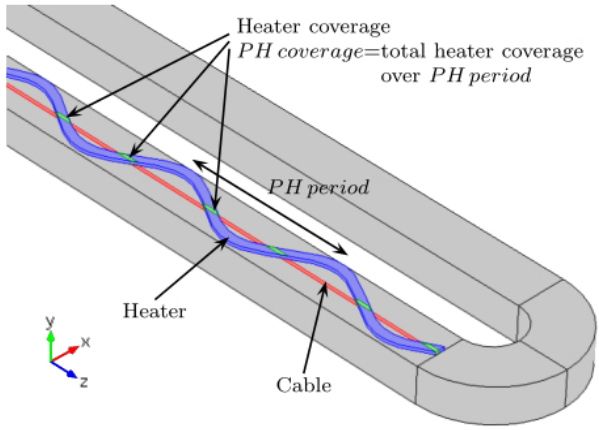

Figure 1: A schematic view showing how generic heater geometry can be expressed in terms of periodical heater coverage at different turns.

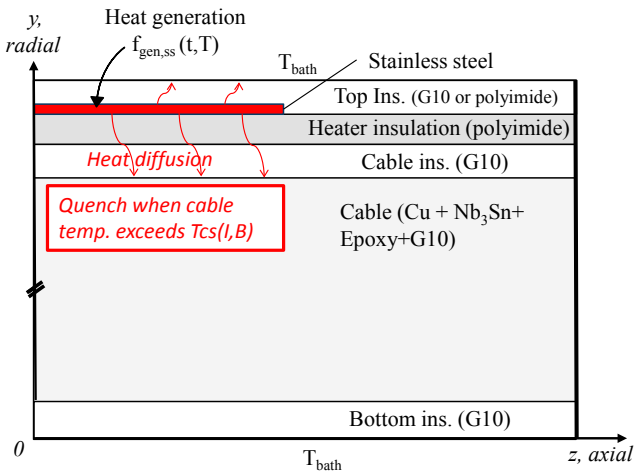

Figure 2: Thermal model for half period of the protection heater geometry, representing the longitudinal and radial (through wide side of the cable) thermal transport in one coil turn.

The material thermal properties are functions of local temperature and magnetic field. The copper properties and epoxy specific heat are from [7] (with linear extrapolation to 0 J/kg/K at 0 K for epoxy specific heat below 4.4 K), $Nb_3Sn$ specific heat is a fit from [8], G10, polyimide (Kapton) and stainless steel properties are from [9] (with extrapolation for Kapton thermal conductivity below 4.3 K [8], and stainless steel specific heat below 5 K [10]). The stainless steel resistivity is based on [11].

The cable is a homogeneous block with properties averaged over its constituents (copper, $Nb_3Sn$, epoxy and/or G10) volume. Thermal conductivities of $Nb_3Sn$ and epoxy are assumed negligible relative to that of copper. By default, the magnetic field in the cable cross-section is uniform, and it is an input parameter. The model allows also simulating variable field profile across the cable. In that case the current sharing temperature, $T_{cs}$, varies at different cable location, and the material thermal properties are based on the average field.

*Quench delay determination*

The simulation begins with the powering of the heaters and the quench delay is defined once the cable temperature exceeds $T_{cs}(B,I)$, i.e. the temperature at which the current in the cable is equal to the (temperature and magnetic field dependent) critical current. The model offers two possibilities for fitting the critical surface, Godeke [12] [13] and Summers [14].

*Numerical solution*

The numerical solution is based on the thermal network method [15] with explicit finite difference discretization scheme [16] and adaptive time stepping. Several elements in each layer are needed to guarantee numerical stability and accuracy. The segments size is an input parameter.
The correct implementation of the equations was verified by comparison with analytical solution of a case in 1-D heat conduction in an insulated slab with steady surface heat flux and constant and uniform material properties.

## SIMULATION OF THE HQ HEATER IN THE HQ01 QUADRUPOLE

As the first study case, the model is applied to the LARP HQ magnet, which is a 1-m-long 120-mm-aperture quadrupole based on cosθ geometry with two layers [4]. The outer layer heater implemented in the coils is modeled, and the simulated heater delays are compared with experimental data from the HQ01e tests [5]. Then, the impact of individual heater design parameters on the quench delay is examined using a parametric analysis. The used coil parameters are shown in Table 1, and the field map in Fig. 3. In the next sections the used parameters for both studies are detailed.

*Simulation of the HQ heater geometry*

The HQ outer layer heater has a wavy shape, providing partial coverage at several turns. One period of the geometry is shown in Fig. 4. It shows that the heater coverage increases from about 2 cm to 7 cm in approximately 1 cm steps in turns 2nd to 7th (counted from the outer layer (OL) pole).

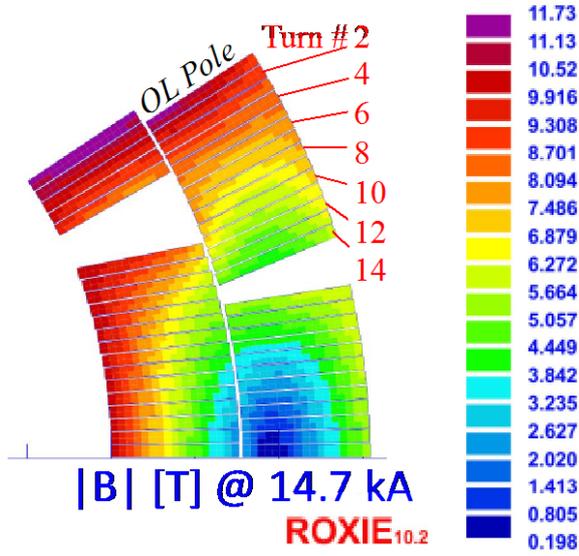

Figure 3: HQ01 field map and notation of the turn count from outer layer pole.

Table 1: Simulation parameters for the LARP HQ simulation

| Parameter | HQ01 (Coil 9) |
|---|---|
| SSL@ 1.9 K (kA) | 19.31 |
| SSL @ 4.4 K (kA) | 17.52 |
| $B_{peak}$, at $I$ [18] | $0.00127 \times I^{0.9505}$ |
| #strands | 35 |
| Copper RRR | 190 |
| Strand Cu/SC | 1.05 |
| Cable voids | 12% epoxy |
| Cable width (mm) | 15.00 |
| Cable ins. (mm) | 0.090 (G10) |
| Bottom ins. (mm) | 0.708 (G10) |
| Top ins. (mm) | 0.30 (G10) |
| Stainless steel (mm) | 0.025 |
| PH ins. Kapton (mm) | 0.0254 |
| Strip path (mm) | 2220.0 |
| Strip width (mm) | 11.0 |

After the 7$^{th}$ turn, the continuous heater coverage is smaller than 7 cm. As the heater coverage increases while moving away from the pole, the magnetic field decreases (Fig. 3). Higher field and longer coverage are assumed to compete in reducing the delay, so the location of the first heater-induced quench is not obvious. In the experiment the first quench can be located between turns 2 and 14 based on voltage tap signals, but it is not known in more detail. Here it is assumed to occur in one of the turns from 2$^{nd}$ to 7$^{th}$ and the heater delay is simulated at each of these turns. The shortest quench delay among the modelled turns is chosen for the comparison with experimental data.

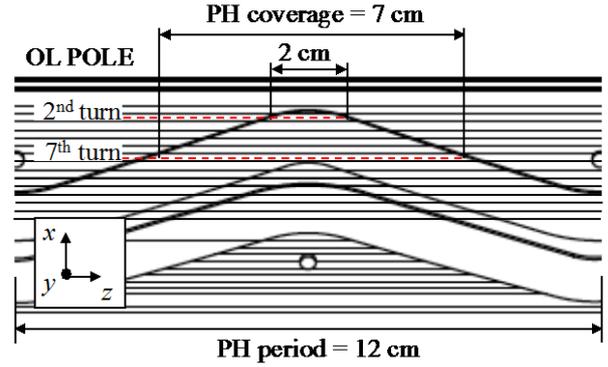

Figure 4: One period of the HQ01 heater on the coil outer surface. The *PH coverage* (the length of the cable continuously covered by the heater at each turn) in one *PH period* (periodical heater geometry) is shown for the 2$^{nd}$ and 7$^{th}$ turn.

The magnetic field strength is calculated at the coil outer surface (using Cobham Vector field Opera-2D [17]), which is the location closest to the protection heater and this value is used for the whole turn. The fields in turns from 2$^{nd}$ to 7$^{th}$ normalized to the magnet peak field at a given current are respectively 0.75, 0.74, 0.72, 0.70, 0.69 and 0.66.

The Nb$_3$Sn critical surface is calculated using Godeke fit with parameters from HQ coil 9 extracted strand measurements [18]. The calculated $T_{cs}$ varies from 14.2 to 14.4 K at 5 kA and from 9.6 to 10.4 K at 14 kA. The heater power is defined by 230 V over the 2220 mm long strip, giving $J_{ss}$ = 210 A/mm$^2$, which gives a heater power $P_{PH}(0)$ about 50 W/cm$^2$. The current decays according to a time constant of 40 ms (defined from the measured current decay profile).

*Parametric study*

In the parametric study, we modeled the outer layer 2$^{nd}$ turn at 1.9 K, and magnet current 80% of the short sample limit (15400 A). The computed conductor field is 9.1 T, and $T_{cs}$ is 8.9 K.

The varied parameters are the heater power, the Kapton thickness, and the heater coverage. If not otherwise mentioned, in the parametric analysis the heater power $P_{PH}$ is 50 W/cm$^2$ and constant (step function), the heater covers the whole turn, and the Kapton thickness is 0.025 mm.

## HQ01 SIMULATION RESULTS

*Comparison with experimental data*

The HQ heater simulation at different turns shows that the delays increase from about 5 to 40 ms when decreasing the magnet current from 80% of short sample limit to 20% (see Fig. 5). The case with infinite heater coverage (1D), at magnet peak field ($B/B_{peak}$ = 1.0) is also shown, and as expected, the delays converge to that when increasing heater coverage or field fraction. The variation between the turns is larger at lower current and the turn that quenches first depends on the current.

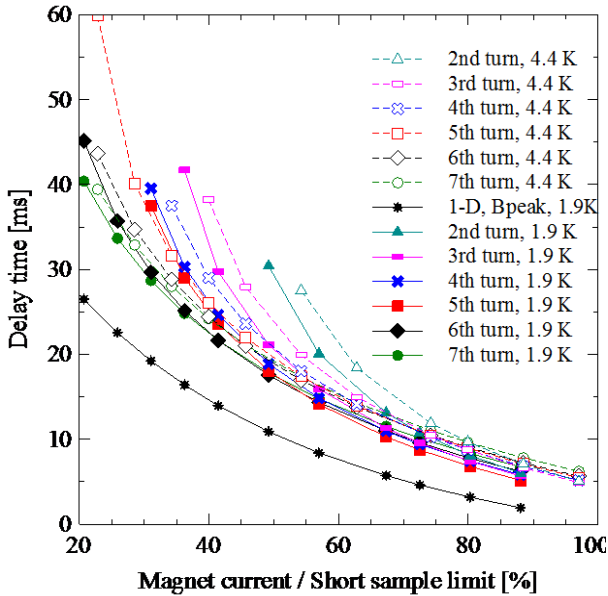

Figure 5: HQ heater delays simulated at several outer layer coil turns. The solid lines represent operation at 1.9 K and dashed lines at 4.4 K.

The simulation agrees with experimental data within 20%, as shown in Fig. 6, where the shortest delays at each current are plotted together with the experimental data. The impact of the operation temperature on the delays is only a few percent in both the simulation and experiment. Excluding the longest simulated delay times (where the heat diffusion away from the hotspot plays a larger role), this difference is approximately proportional to the difference in the energy margins to quench (i.e., integration of the cable heat capacity from $T_{bath}$ to $T_{cs}$) at each current.

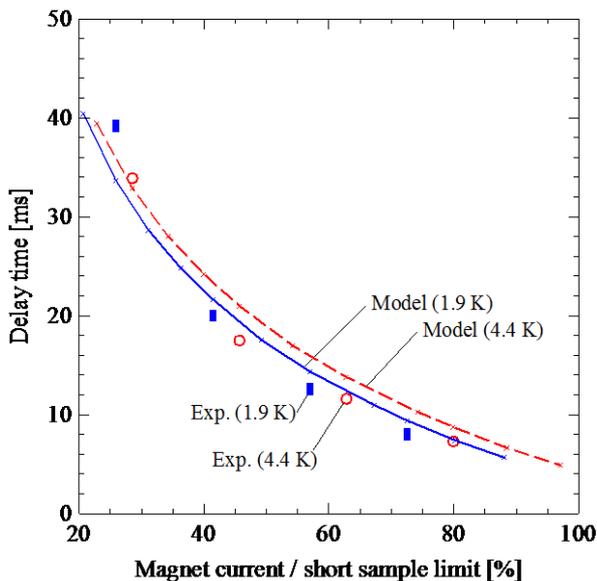

Figure 6: Modeled and experimental (Exp.) HQ heater delays at 1.9 and 4.4 K versus normalized magnet current.

*Heater delay vs. heater power*

As expected, larger heater power reduces the simulated delays, as shown in Fig. 7. Saturation is visible around 30 W/cm$^2$. Increasing the power further has only a small effect on the delay. The curve shape is consistent with experiments [1].

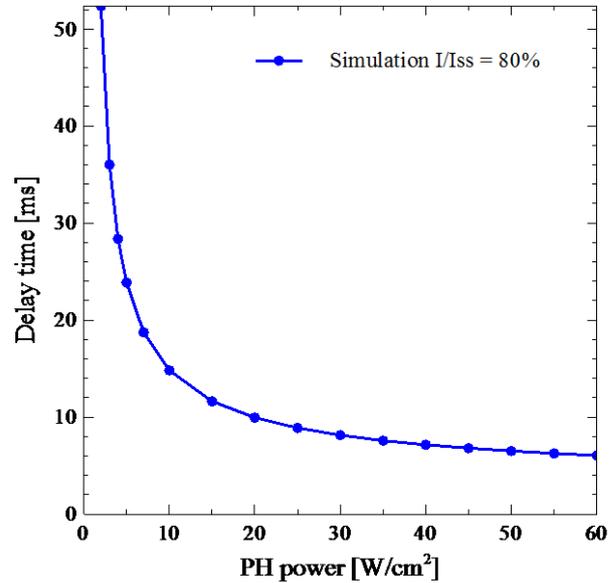

Figure 7: Heater delay time vs. heater peak power. The heater power is a step function in time.

*Delay time vs. insulation thickness*

The increase in the simulated delay when increasing the polyimide thickness is shown in Fig. 8. The delay approximately doubles when the thickness is increased from 0.025 mm to 0.076 mm. Comparison of experimental data from HQ01e (0.025 mm Kapton), and from HQ coil 15 (0.076 mm Kapton), which was tested in the HQM04 mirror structure, shows an increase in the experimental delay approximately 130%, in agreement with the simulated value.

*Delay time vs. heater geometry*

The simulation shows that longer heater coverage leads to shorter delays – up to saturation around 20 mm, when the delay approaches 7 ms indicating a local 1-D heat transfer (fully covered cable) (see Fig. 9). At coverage of 5 mm, the delay is more than doubled. Variation of the period between 50 and 180 mm changed the result less than 5% with respect to the reference case with 120 mm long period.

Longer delay for the same short sample fraction was also found in the LARP LQ magnet, which had shorter heater coverage than HQ [3].

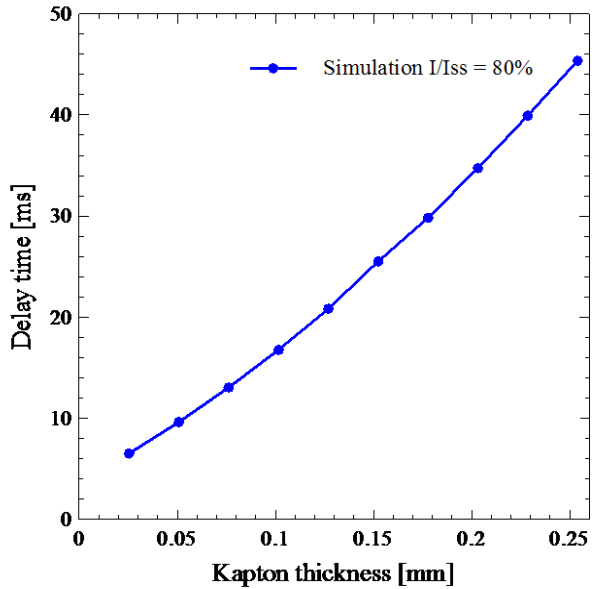

Figure 8: Heater delay vs. Kapton thickness.

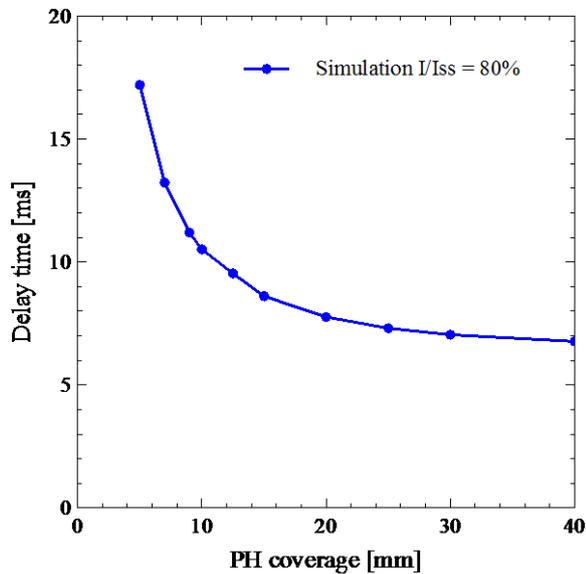

Figure 9: Heater delay time vs. length of the covered cable segment.

## SIMULATION OF THE HEATER IN THE 11 T DIPOLE

Contrary to HQ, the heater strips in the 2-m long so-called 11 T dipole model (MBHSP01) are straight strips parallel to the magnet axis. Therefore the problem is essentially in 1-D, when the heat transfer between the coil turns is neglected.

The heaters have been tested with Kapton thickness of 0.076 mm and 0.203 mm, i.e. much larger than HQ. Therefore the simulation of the 11 T dipole heaters allows applying the model to a quite different regime. The adhesive (0.038 mm) that is used to glue the stainless steel heater to the Kapton has been neglected in the simulation. The 11 T dipole magnet development and test are described in [6] and the protection heater experiments in [19].

*Heater power*

On the outer surface of each coil half, two straight heater strips form a U-shape and are connected in series (see Fig. 10 [19]). The strip closer to the central pole piece (high field region) is 26 mm wide, and the strip closer to the magnetic midplane (low field region) is 21 mm wide. Two U-shapes are connected in parallel and powered by a capacitor discharge in one Heater Firing Unit (HFU). The capacitance is 9.6 mF and the measured cold resistance of the circuit is 2.6 Ω, giving RC-time constant of 25 ms.

The calculated heater current is 77 A for a voltage of 400 V. Using equation (1) and multiplying by the stainless steel thickness we get a peak power of about 17 W/cm$^2$ in the high field heater and 27 W/cm$^2$ in the low field heater, using the stainless steel 304 resistivity (490 nΩm @ 4.2 K). The calculated resistance of both heaters together is 3.4 Ω. However, the measured resistance of the U-shaped heater is 20% larger, 4.2 Ω. Partial explanation is that the heaters in the 11 T dipole are based on stainless steel 316 L, which has about 5% higher electrical resistivity. Assuming that the measurement gives the correct resistivity (and for example the connection in between the strips or irregularities in the heater shape do not impact), the heater power is 20 W/cm$^2$ in the high field heater, and 31 W/cm$^2$ in the low field heater. In the simulation we use the average of these: 18.5 W/cm$^2$ in the high field and 29 W/cm$^2$ in the low field region.

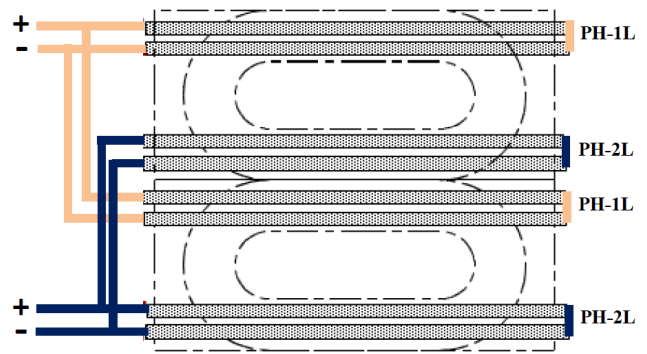

Figure 10: 11 T dipole heater connection scheme. PH-1L and PH-2L refer to heater insluation thickness with 1 or 2 layers of Kapton [19].

*Magnetic field and cable properties*

Under each heater, the first quench is expected to initiate at the coil turn that has the highest magnetic field. The high field and low field heater were considered separately, and the delay was simulated in the turns #2 and #19 from the outer layer pole (see Fig. 11 [20]).

The choice of field value for the turn #2 is not straight forward because the field on the coil outer diameter (OD) is only 78% of the maximum field of that conductor (see Fig. 10). We therefore considered three cases. In Case 1, field was taken at the coil outer surface (65% of the magnet peak field). In Case 2, field was taken as the maximum field in the conductor (82% of the magnet peak field). And, in Case 3, the field profile varies across the conductor (1-D projection of the 2-D field map in the cable cross-section). In the turn #19, the field at the coil surface is the same as the cable maximum field (42% of the magnet peak field), so simulations were done only for this field value.

The HQ simulation corresponds to the Case 2. The field location in the 11 T simulation is more critical for two reasons: First, in HQ the cable outermost field was 87-95% of the maximum field. Second, in 11 T the expected delays are longer due to smaller heater power and thicker insulation between the heater and cable. The longer delays increase the impact of all factors, including the field. One should keep in mind that while tuning the field location may be useful for finding the best expectation for the experimental results, it may give a false sense of accuracy because the anisotropic cable internal structure (strands' paths) is still not modelled.

The critical surface is based on the Summers fit, using $B_{c20}$ = 24.8 T, $T_{c0}$ = 16.5 K, C = 9.08×10$^3$. Other simulation parameters are summarized in Table 2.

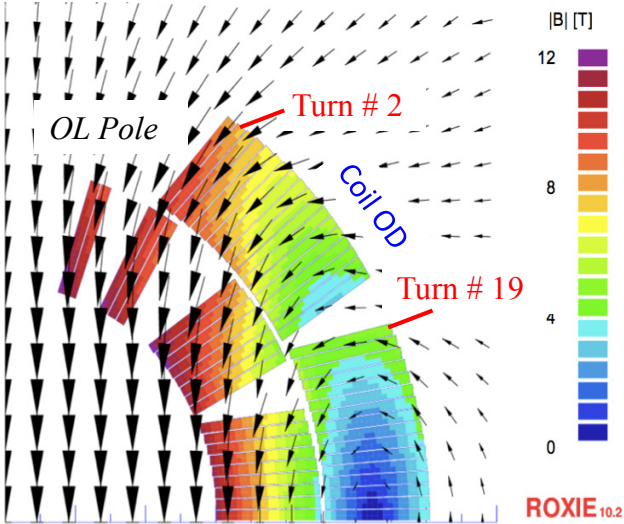

Figure 11: 11 T dipole field map [20].

Table 2. Simulation parameters for CERN-FNAL 11 T dipole simulation

| Parameter | 11 T |
|---|---|
| SSL @ 1.9 K (kA) | 15.4 |
| SSL @ 4.4 K (kA) | 13.8 |
| $B_{peak}$, at I [18] | $0.0023 \times I^{0.9062}$ |
| #strands | 40 |
| Copper RRR | 100 |
| Strand Cu/SC | 1.13 |
| Cable voids | 6.5% epoxy, 6.5% G10 |
| Cable width (mm) | 14.88 |
| Cable ins. (mm) | 0.200* (G10) |
| Bottom ins. (mm) | 0.706 (G10) |
| Top ins. (mm) | 0.64 (Kapton) |
| Stainless steel (mm) | 0.025 |
| PH ins. Kapton (mm) | 0.076 / 0.203 |
| Strip path (mm) | 2100.0 |
| Strip width (mm) | 26.1 |

* This refers to the insulation between the bare cable and the polyimide of the PH insulation. In the 11 T magnet 0.2 mm includes a 0.1 mm glass sheet that is impregnated on the coil surface.

## 11 T DIPOLE SIMULATION RESULTS

The heater delays were measured between 40 and 60% of SSL. Simulations in general show a good agreement with results, giving (i) much longer delays for thicker polyimide and (ii) the correct slope of delay increase at lower currents. At 80% of SSL at 1.9 K the heater delay is expected to be about 55 ms with 0.203 mm Kapton, and 25 ms with 0.076 mm Kapton. The 20% predicted increase in the simulated delays from 1.9 to 4.5 K is not seen in the experiment.

The simulated delays at 1.9 K agree the best with the experimental data for the high field heater when the utilized field was the maximum in the cable (Case 2). The agreement is within 20% for both thicknesses when above 50% of SSL at 1.9 K. The delays using the realistic field profile (Case 3) are about 10-30% longer than the delays with the maximum field. When the field is taken at the coil OD (Case 1), the delay is at least 60% longer than with the maximum field. The delays under the low field heater were about 50-150% longer than the shortest delays under the high field heater. Figures 12 and 13 show the results in the Cases 2 (Bmax) and 3 (Bprof) of the high field heater.

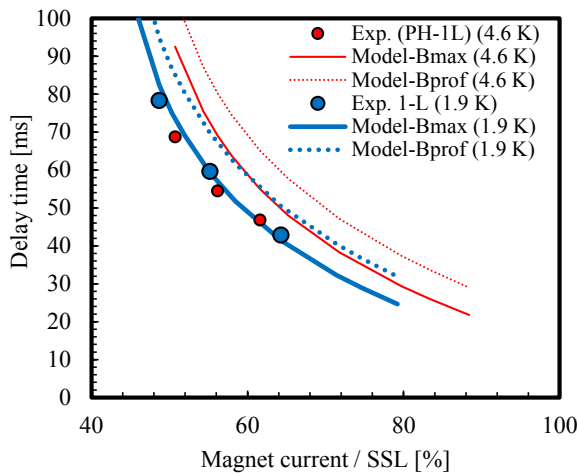

Figure 12: Heater delays in the 11 T dipole, simulated and measured, for the 0.076 mm Kapton thickness.

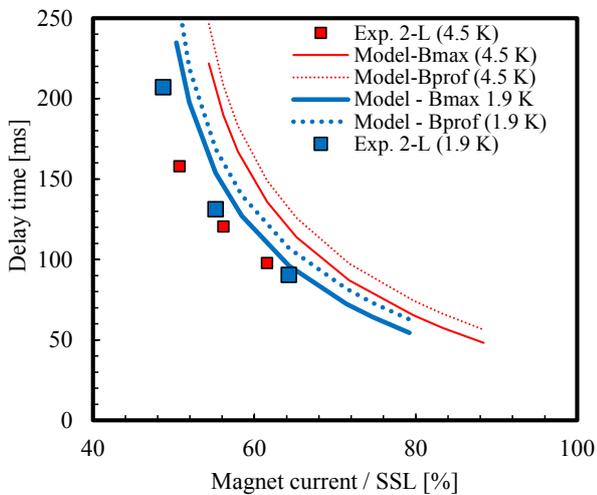

Figure 13: Heater delays in the 11 T dipole, simulated and measured, for the 0.203 mm Kapton thickness.

## CONCLUSION

A computational tool based on a two-dimensional heat conduction model is developed to calculate the protection heater delay time to induce a quench as a function of a large amount of parameters, which include cable properties, magnet operation conditions, and heater geometry, powering and insulation scheme.

The modeling tool is applied to simulate heater delays in the LARP $Nb_3Sn$ quadrupole magnet called HQ01e and in the FNAL-CERN $Nb_3Sn$ dipole magnet called 11 T. The agreement between the simulation, which does not use any free parameters, and experimental data is within 20% in most cases. A parametric analysis using the HQ01e data showed the heater delay dependence on heater power, polyimide thickness and heater geometry.

This relatively simple modeling approach can be useful in understanding the effect of various parameters on the quench delay time, which is important for optimizing the heater design for future high-field $Nb_3Sn$ accelerator magnets.

## ACKNOWLEDGMENT

This work has been supported by the U. S. Department of Energy, under Contract No. DE-AC02-05CH11231. Finalizing this paper was supported by Stability Analysis of Superconducting Hybrid magnets (Academy of Finland, #250652). The authors wish to thank Hugo Bajas and Jerome Feuvrier (CERN) for the heater delay measurements in HQ01e. We thank Fred Nobrega (FNAL), Bernhard Auchmann (CERN) and Emanuela Barzi (FNAL) for providing information for the 11 T simulation. I also gratefully acknowledge the support and useful advice on the work from Ezio Todesco (CERN) and Antti Stenvall (Tampere University of Technology), who also provided Figure 1. I have also received helpful input from Maxim Marchevsky, and Matthijs Mentink (LBNL). I also thank Ray Hafalia (LBNL) for grammar corrections in an earlier draft of the paper (the new mistakes in the present version are not his fault).